\documentclass[9pt, conference]{IEEEtran}
\IEEEoverridecommandlockouts
\usepackage{cite}
\usepackage{amsmath, amssymb, amsfonts}
\usepackage{algorithmic}
\usepackage{graphicx} 
\usepackage{textcomp}
\usepackage{xcolor}
\usepackage{comment}
\usepackage{colortbl}
\usepackage{dhucs} 
\usepackage{booktabs} 
\usepackage{multirow}
\usepackage{setspace} 
\usepackage{makecell} 
\usepackage{enumitem} 
\usepackage{nicematrix} 
\usepackage{threeparttable} 
\usepackage{dblfloatfix} 
\usepackage[ruled,vlined,algo2e]{algorithm2e}
\usepackage{lineno}
\usepackage{relsize}
\usepackage{tabularx}
\usepackage{diagbox}
\usepackage{hhline}

\newcommand\refFigure[1]{Fig.~\ref{#1}}
\newcommand\refTable[1]{Table~\ref{#1}}

\newcommand\refAlgo[1]{Algorithm~\ref{#1}}

\newcommand\korean[1]{}
\newcommand\blind[1]{XXXX}

\def\useshortref{0}
\def\usemiddleref{0}
\if\useshortref 1

\else

\if\usemiddleref 1

\else

\fi
\fi

\if\useshortref 1

\else

\if\usemiddleref 1

\else

\fi
\fi

\def\BibTeX{{\rm B\kern-.05em{\sc i\kern-.025em b}\kern-.08em
    T\kern-.1667em\lower.7ex\hbox{E}\kern-.125emX}}

\IEEEpubid{\makebox[\columnwidth]{ This paper will appear in the proceedings of ISLPED 2025. 
\hfill{ \hspace{\columnsep} \makebox[\columnwidth]{ }}}}

\begin{document}

\title{ASAP-FE: Energy-Efficient Feature Extraction Enabling Multi-Channel Keyword Spotting on Edge Processors\\
\thanks{This work was supported in part by Institute of Information \& communications Technology Planning \& Evaluation (IITP) grants funded by the Korea government (MSIT) No. RS-2023-00277060 (Development of open edge AI SoC hardware and software platform) , and No. 2022-0-00957 (Distributed On-Chip Memory Processor Model PIM Semiconductor Technology Development for Edge Applications), and in part by the US National Science Foundation.}
\thanks{$^{*}$Jongin Choi and Jina Park contributed equally to this work.}
\thanks{$^{\#} Corresponding~authors$: Woojoo Lee ($space@cau.ac.kr$)}}
\author{
Jongin Choi$^{\dag *}$, Jina Park$^{\dag *}$, Woojoo Lee$^{\dag \#}$, Jae-Jin Lee$^{\ddag}$, and Massoud Pedram$^{\S}$\\
$^{\dag}$ Department of Intelligent Semiconductor Engineering, Chung-Ang University, Korea \\
$^{\ddag}$ AI SoC Research Division, Electronics and Telecommunications Research Institute (ETRI), Korea \\
$^{\S}$ Department of Electrical and Computer Engineering, University of Southern California, USA
}

\maketitle

\begin{abstract}
Multi-channel keyword spotting (KWS) has become crucial for voice-based applications in edge environments. However, its substantial computational and energy requirements pose significant challenges. We introduce \emph{ASAP-FE} (\textit{Agile Sparsity-Aware Parallelized-Feature Extractor}), a hardware-oriented front-end designed to address these challenges. Our framework incorporates three key innovations: (1) Half-overlapped Infinite Impulse Response (IIR) Framing: This reduces redundant data by approximately 25\% while maintaining essential phoneme transition cues. (2) Sparsity-aware Data Reduction: We exploit frame-level sparsity to achieve an additional 50\% data reduction by combining frame skipping with stride-based filtering. (3) Dynamic Parallel Processing: We introduce a parameterizable filter cluster and a priority-based scheduling algorithm that allows parallel execution of IIR filtering tasks, reducing latency and optimizing energy efficiency. ASAP-FE is implemented with various filter cluster sizes on edge processors, with functionality verified on FPGA prototypes and designs synthesized at 45 nm. Experimental results using TC-ResNet8, DS-CNN, and KWT-1 demonstrate that ASAP-FE reduces the average workload by 62.73\% while supporting real-time processing for up to 32 channels. Compared to a conventional fully overlapped baseline, ASAP-FE achieves less than a 1\% accuracy drop (e.g., 96.22\% vs. 97.13\% for DS-CNN), which is well within acceptable limits for edge AI. By adjusting the number of filter modules, our design optimizes the trade-off between performance and energy, with 15 parallel filters providing optimal performance for up to 25 channels. Overall, ASAP-FE offers a practical and efficient solution for multi-channel KWS on energy-constrained edge devices.
\end{abstract}

\vspace{2pt}
\section{Introduction}
 
Keyword Spotting (KWS) is a core human-computer interface that activates a system immediately upon detecting pre-trained keywords. 
As voice-based devices continue to spread rapidly, KWS has found increasing relevance in various domains such as the IoT and security systems \cite{Sudharsan:WF-IoT2021, Oliver:INFOTECH2023}, and there is growing interest in leveraging KWS for unattended monitoring environments. 
In particular, \emph{Multi-channel Intelligent Surveillance Systems}, which integrate multiple audio channels to enhance real-time monitoring capabilities, require KWS to quickly and accurately handle large-scale audio data simultaneously gathered under various conditions \cite{Zhang:ICASSP2023, Su:AI2022, Li:TCSVT2025}. 
However, most KWS technologies have been designed for mono-channel inputs, leaving significant challenges in managing the massive data load, noise adaptation, and real-time (latency) requirements presented by multi-channel inputs.

In edge environments, \emph{low power and high energy efficiency} are crucial design objectives~\cite{Han:ISLPED19,Jina:ISLPED23,Kwak:ISLPED24,Jeon:DAC25}. However, the computational burden increases exponentially with the number of channels, making it increasingly challenging to support real-time multi-channel keyword spotting (KWS). A typical KWS pipeline consists of a front-end (Feature Extraction; FE) and a back-end (Classifier). Although lightweight classifier models such as DS-CNN \cite{Zhang:HelloEdge} and TC-ResNet8 \cite{Choi:TCResNet} can partially alleviate performance degradation in multi-channel scenarios, the front-end, long recognized as the primary bottleneck in single-channel KWS \cite{Shan:JSSC2021} faces particularly severe computational demands. 
Furthermore, addressing various types of noise during the signal acquisition stage, rather than increasing classifier complexity, is much more suitable for energy-constrained edge devices. Consequently, the need for \emph{efficient processing of multiple audio channels in the FE} has become increasingly critical.

\begin{figure}[t]
    \centering
    \vskip -2pt
    \includegraphics[width=0.98\columnwidth]{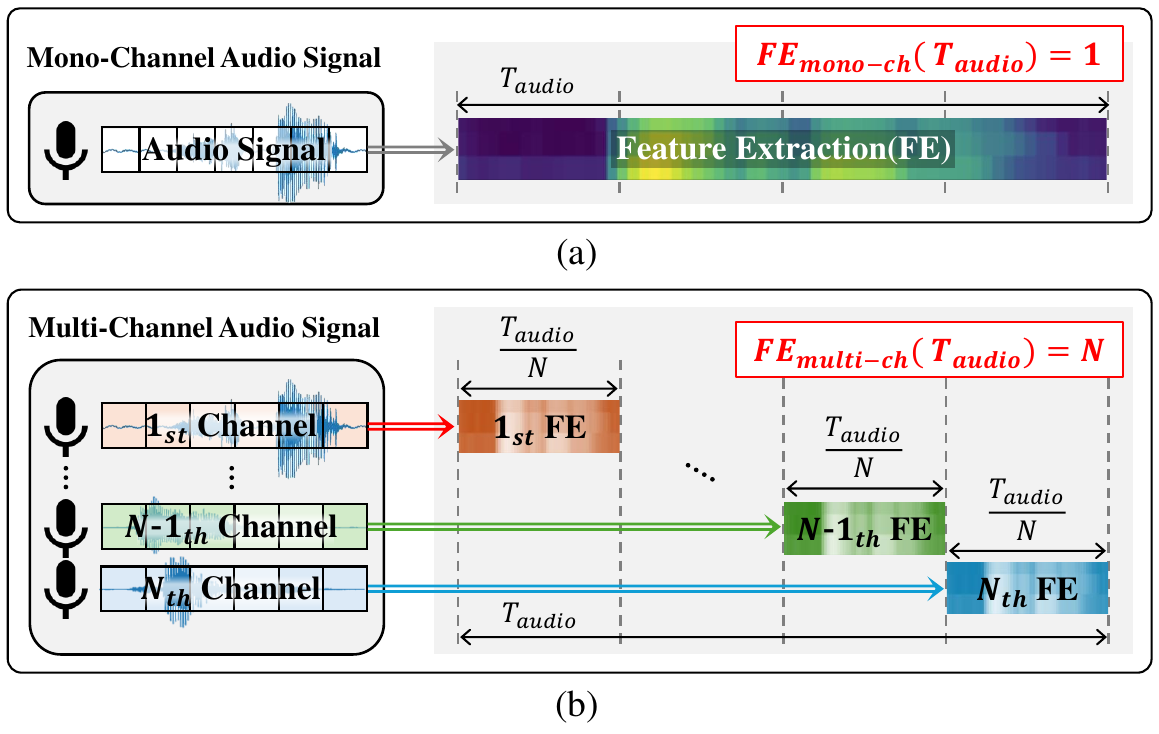}
    \vskip -6pt
    \caption{(a) Mono-channel IIR-FBank: only one audio stream per $T_{audio}$, and (b) Shorter FE for multi-channel KWS: multiple channels within the same $T_{audio}$.}
    \label{fig:intro1}
\end{figure}

Previous KWS FE research has taken both analog and digital approaches. 
Analog methods, such as ring-oscillator bandpass filters (BPFs) \cite{Kim:JSSC2022} and transconductance-capacitor ($g_mC$) BPFs \cite{Wang:ISSCC2021}, typically achieve around 90\% accuracy or lower and are highly sensitive to noise, making them unsuitable for multi-channel KWS with complex noise conditions. 
Digital approaches, which typically offer higher accuracy than analog methods, commonly employ Mel-Frequency Cepstral Coefficients (MFCC) based on FFT (Fast Fourier Transform)-based time-frequency transforms. 
However, this process imposes high overhead \cite{Shan:JSSC2021, Shan:ISSCC2020, Li:JSSC2024, Wu:Electronics2023} on edge processors. 
Although Infinite Impulse Response (IIR) filter bank-based FE (referred to as \textit{IIR-FBank}) \cite{Yu:TCASII2024, Chen:ISCAS2023} has been proposed to address these limitations, its recursive nature still restricts processing to a single input stream at a time, as illustrated in \refFigure{fig:intro1}(a), thus limiting straightforward application to multi-channel setups. 
Consequently, as shown in \refFigure{fig:intro1}(b), a shorter and more efficient FE architecture is needed to process multiple channels in parallel or fast sequential mode within the same $T_{audio}$.

\begin{figure}
    \centering
    \vskip -12pt
    \includegraphics[width=1\columnwidth]{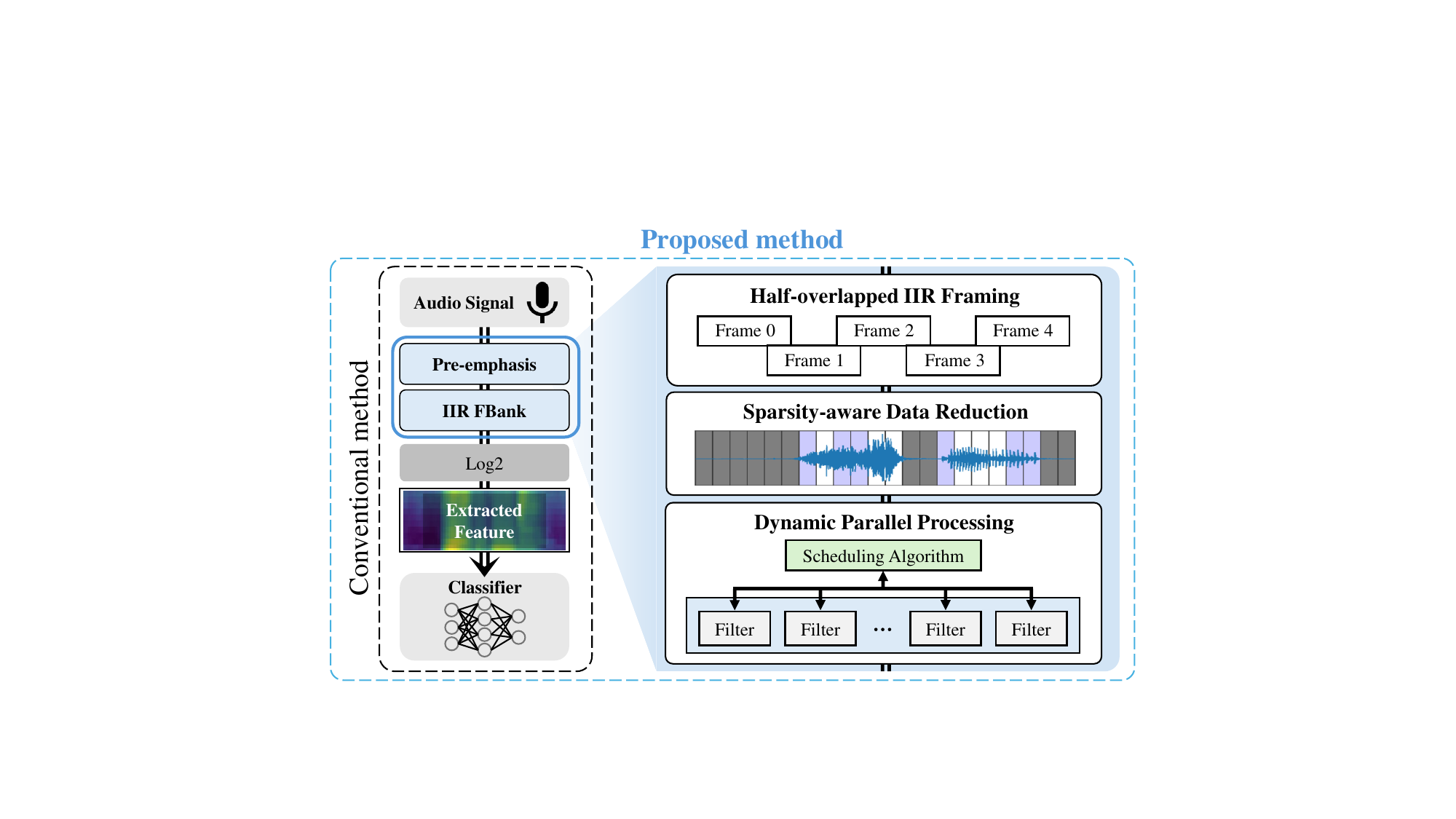}
    \vskip -4pt
    \caption{Conceptual overview of the proposed ASAP-FE.}
    \label{fig:intro2}
     \vskip 2pt
\end{figure}

As \refFigure{fig:intro2} shows, apart from the \textit{log2} operation required to adapt the features to the human auditory characteristics, most computations in an IIR-FBank are spent on IIR filtering itself \cite{Yu:TCASII2024}. 
Motivated by this observation, our work targets two main objectives for building a multi-channel KWS solution: \emph{(1) reducing the amount of data to be filtered and (2) exploiting parallel filtering to improve execution time and energy efficiency}. Specifically, we propose three key techniques:
\begin{enumerate}
    \item \textbf{Half-overlapped IIR Framing}:  
    This method enables parallel operation by segmenting input signals into frames while reducing frame overlap, preserving crucial temporal speech cues, and reducing redundant data by approximately 25\%.
    \item \textbf{Sparsity-aware Data Reduction}:  
    By leveraging frame-level sparsity to skip or rapidly process less significant frames, we achieve an additional 50\% data reduction.
    \item \textbf{Dynamic Parallel Processing}:  
    We design multiple parallel filter modules and an optimization-driven scheduling algorithm to handle input data simultaneously, minimizing processing latency. We also determine the optimal number of filter modules for maximizing energy efficiency.
\end{enumerate}

By integrating these techniques, we develop \emph{Agile Sparsity-Aware Parallelized-Feature Extractor (ASAP-FE)}, a hardware architecture shown in \refFigure{fig:intro2}.
We implement edge processors incorporating ASAP-FE with various filter module counts at the full RT level (RTL), verifying functionality and performance on an FPGA running at 50\,MHz.
We then synthesize the designs in a 45\,nm technology node and perform power simulations to evaluate their energy efficiencies.
Experiments on three representative KWS models—TC-ResNet8, DS-CNN, and KWT-1~\cite{Berg:Interspeech2021}—demonstrate that the developed processors achieve accuracies of 95.43\%, 96.22\%, and 96.48\%, respectively, compared to conventional results of 96.35\%, 97.13\%, and 97.28\%.
This amounts to less than 1\% accuracy drop in each case, remaining within commonly acceptable limits~\cite{Xiao:ISSCC2024, Xiao:ISCAS2024}.

Meanwhile, half-overlapped IIR framing and sparsity-aware data reduction reduce the overall workload by 62.73\%. 
Further, with dynamic parallel processing, the prototype processor can handle up to 32 channels within the same \emph{$T_{audio}$} as a single-channel KWS. 
In addition, configuring 15 parallel filter modules minimizes energy consumption and yields optimal efficiency for real-time processing of up to 25 channels. 
These results demonstrate that ASAP-FE is an effective solution for meeting performance and energy constraints in multi-channel KWS on edge devices.

\section{Half-overlapped IIR Framing} \label{sec:framing_methods}

\normalsize
An IIR-FBank extracts mel-frequency band information by employing multiple IIR bandpass filters (BPFs) that densely cover low frequencies and more sparsely cover high frequencies, mirroring human auditory perception. 
Unlike conventional MFCC-based systems, an IIR-FBank does not require FFT operations and thus offers a lower computational cost when implemented in hardware. 
In general, an IIR BPF can be expressed as

\begin{figure}[t]
\vskip -4pt
    \centering
    \includegraphics[width=1\columnwidth]{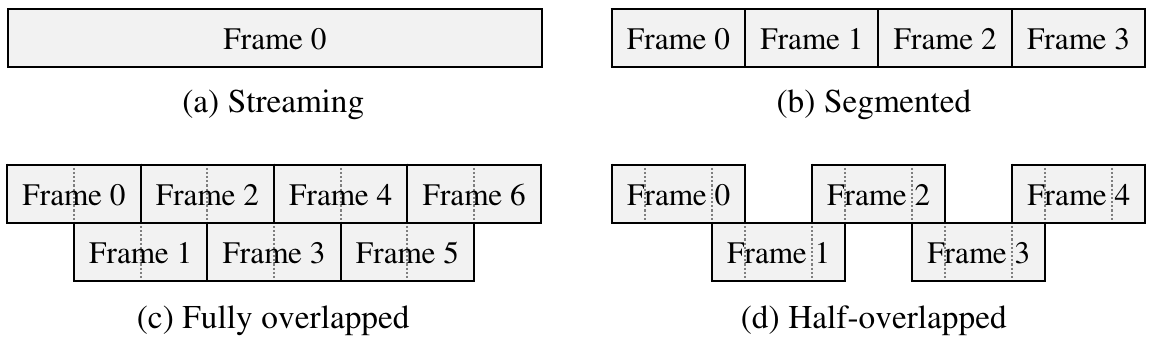}
    \vskip -4pt
    \caption{Examples of streaming and framing methods.}
    \label{fig:framing}
    \vskip 2pt
\end{figure}

\vspace{-8pt}
\begingroup
\footnotesize
\begin{align}
\label{eq:eq1}
    y[n] = -\sum_{k=1}^{N} a_k \, y[n - k] + \sum_{k=0}^{M} b_k \, x[n - k],
\end{align}
\endgroup
where \( x[n] \) and \( y[n] \) denote the filter input and output at discrete-time index \( n \); \( a_k \) and \( b_k \) are the feedback and feedforward coefficients, respectively; and \( N \) and \( M \) represent the orders of the feedback and feedforward paths of the filter.
As in \eqref{eq:eq1}, an IIR BPF processes input samples and output values throughout the length of the speech signal in the time domain, continually computing the current output from previous outputs. 
Consequently, the conventional implementation follows the \emph{streaming} approach depicted in \refFigure{fig:framing}(a). 
A streaming IIR filter achieves accuracy comparable to conventional MFCC-based feature extraction~\cite{Yu:TCASII2024}, rendering it sufficiently accurate for KWS.

However, streaming-based IIR filtering cannot exploit frame-by-frame parallelism, limiting the potential for higher throughput. 
Hence, introducing a framing strategy becomes crucial under real-time constraints (as in multi-channel KWS). 
Although a \emph{segmented} framing approach for IIR-based KWS has been explored \cite{Chen:ISCAS2023} (\refFigure{fig:framing}(b)), it does not implement parallel frame processing. 
Moreover, it omits frame overlap, causing phoneme-transition information to be lost at frame boundaries, thereby reducing accuracy. 
A potential solution is the \emph{fully overlapped} framing method (\refFigure{fig:framing}(c)), which is common in MFCC pipelines \cite{Shan:JSSC2021, Pattanayak:Interspeech2022, Rout:SPECOM2023} (though not yet widely applied in IIR approaches), but fully overlapped framing entails some amount of repeated filtering for each input segment.

\begin{table}[t]
    \caption{Comparison of conventional Feature Extraction Methods and the Proposed Approach.}
    \centering
    \vskip -4pt

\renewcommand{\arraystretch}{1.2} 
\resizebox{1\columnwidth}{!}{%
\renewcommand{\arraystretch}{1.3}
\begin{tabular}{c|cccc}
\Xhline{1pt}
\diagbox[dir=NW,width=3.1cm,height=1.3\line]{}{} & MFCC & \cite{Yu:TCASII2024} & \cite{Chen:ISCAS2023} & \textbf{ASAP-FE} \\  
\Xhline{1pt}
Accuracy(\%) & 96.1 & 96.49 & 92.5 & \textbf{96.48} \\ \hline

Bottleneck & FFT & IIR BPF & IIR BPF & \textbf{IIR BPF} \\ \hline

Framing & Fully overlapped & Streaming & Segmented & \textbf{Half-overlapped} \\ \hline

Parallel-Processing & Feasible & Infeasible & Infeasible & \textbf{Feasible} \\ \hline

\end{tabular}%
}

    \label{table:related}
   \vskip 2pt
\end{table}

\begin{figure}[t]
\vskip -6pt
    \centering
    \hskip -8pt
    \includegraphics[width=0.93\columnwidth]{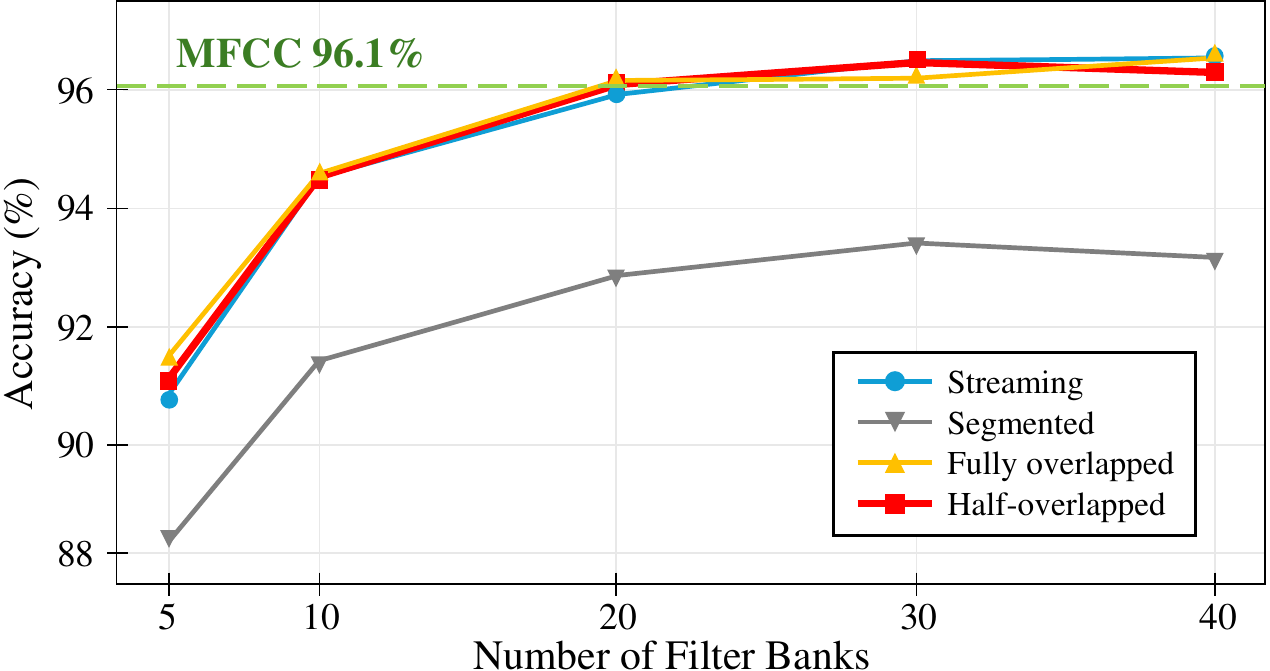}
    \vskip -4pt
    \caption{Impact of different framing methods on KWS accuracy across various filter bank configurations.}
    \label{fig:frame}
\end{figure}

\begin{figure}
\vskip -5pt
\centering
\includegraphics[width=0.98\columnwidth]{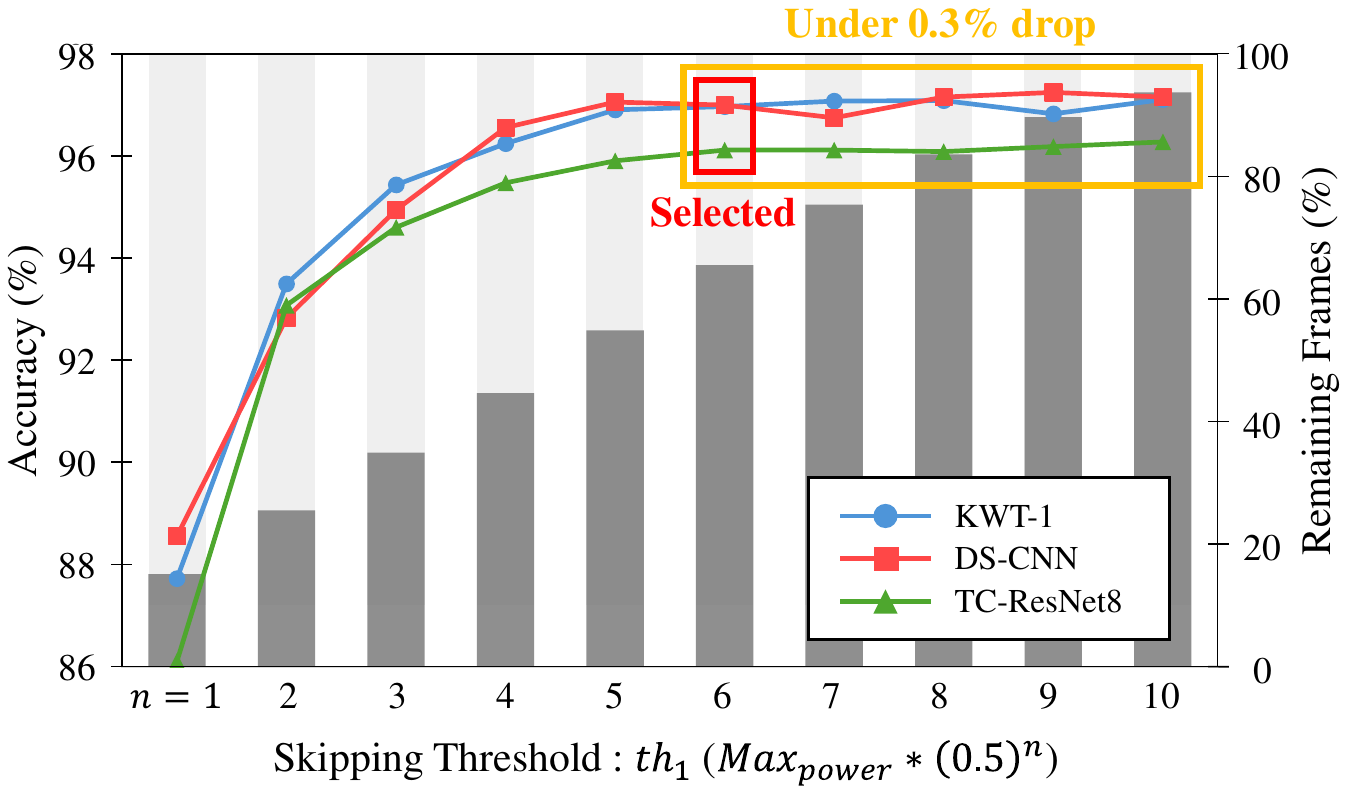}
\vskip -6pt
\caption{Effect of the frame skipping threshold $th_1$ on KWS accuracy and remaining frames.}
\label{fig:skip}
\vskip 2pt
\end{figure}

To address these concerns, we propose the \emph{half-overlapped} framing method shown in \refFigure{fig:framing}(d) for our ASAP-FE design. 
Half-overlapped framing preserves temporal cues via overlapping frames but avoids excessive redundancy. 
Compared to fully overlapped framing, it reduces the overall computation by about 25\% while maintaining accurate phonetic transitions. 
Table~\ref{table:related} compares the accuracy, framing approaches, and parallel processing feasibility of conventional MFCC, previous IIR-based methods \cite{Yu:TCASII2024, Chen:ISCAS2023}, and our proposed half-overlapped framing in ASAP-FE. 
For example, a streaming IIR approach~\cite{Yu:TCASII2024} yields MFCC-level accuracy but lacks parallel framing; a segmented approach loses roughly 4\% accuracy (down to about 92.5\%) due to missing phoneme transitions; and ASAP-FE achieves 96.48\% accuracy by adopting half-overlapped framing, matching existing high-accuracy baselines while enabling frame-level parallelization.

To further investigate the effect of framing strategies on model performance across different numbers of filter banks, we conducted one-keyword spotting experiments under four different framing methods. 
We controlled all conditions except for the number of filter banks. 
We used the Google Speech Commands Dataset (GSCD), which consists of 65,000 1-second audio clips sampled at 16\,kHz, covering 10 target keywords (yes, no, up, down, left, right, on, off, stop, go), plus \emph{silence} and \emph{unknown}, for a total of 12 classes. 
We used random noise augmentation from the GSCD noise set and partitioned the entire dataset into training, validation, and test sets at an 8:1:1 ratio. 
Each audio clip was framed into 16\,ms segments, with mel-scale filter banks reflecting human auditory perception applied for feature extraction. 
Unless otherwise specified, all subsequent experiments in this work use the same dataset and augmentation scheme.

To isolate the framing effect from model-specific influences, we used a fixed and lightweight KWS architecture, TC-ResNet8—trained in PyTorch with Adam~\cite{Kingma:Arxiv2014}, CrossEntropy loss~\cite{Mao:ICML2023}, and a cosine annealing scheduler~\cite{Ilya:ICLR2017}, keeping all hyperparameters constant. 
\refFigure{fig:frame} compares the accuracy of the four framing methods with a baseline that employs 40 MFCC-based filter banks at a reported accuracy of 96.1\%~\cite{Choi:TCResNet}. 
In particular, the segmented approach yields a 3\% or more accuracy drop because phoneme-transition features are lost at frame boundaries. 
In contrast, our half-overlapped method reduces data redundancy by about 25\% relative to full overlap while preserving sufficient temporal detail to match or exceed the baseline accuracy once the filter bank count is above 20. 
These findings confirm that the proposed Half-overlapped IIR Framing is best suited for multi-channel KWS, balancing accuracy with reduced redundancy.

\section{Sparsity-aware Data Reduction} 
Although audio input spans low and high frequencies, many frames contribute little to KWS due to noise or lack of relevant signal \cite{SeoI:ISSCC2023}. 
We can significantly reduce the amount of data to be processed by eliminating or reducing these unnecessary frames. 
To this end, we propose a \emph{Sparsity-aware Data Reduction} technique, which considers the sparsity of each frame to discard irrelevant content while preserving essential KWS information.
 The proposed method comprises two main steps: \textit{Frame Skipping} and \textit{Sparsity-aware Filtering}.

\subsection{Frame Skipping} 
 Each audio input signal is divided into multiple frames, each containing discrete-time samples \(x[n]\). The short-time energy (\textit{STE}) of a frame is defined as the sum of the squared sample values, as expressed: 

\vspace{-12pt}
\begingroup
\footnotesize
\begin{align}
\label{eq:eq2}
\textit{STE} = \sum_{n=0}^{N-1} x[n]^2
\end{align}
\endgroup
Frames with very low \(\textit{STE}\) generally provide little useful information for keyword spotting (KWS), since they are often dominated by noise. Therefore, we adopt a \textit{Frame Skipping} strategy that discards frames whose \(\textit{STE}\) falls below a threshold \(th_1\), thus dedicating more computational resources to frames that more meaningfully contribute to KWS performance.

To determine an optimal \(th_1\) that maximizes frame skipping without incurring significant accuracy loss, we applied half-overlapped framing with 40 mel-scale filter banks~\cite{Zhang:HelloEdge, Choi:TCResNet} and evaluated TC-ResNet8, DS-CNN, and KWT-1 models. 
We defined \(th_1\) as \(0.5^n\) times the maximum \(\textit{STE}\) observed for each audio input, varying \(n \in \{1, 2, \dots, 10\}\). 
\refFigure{fig:skip} shows how accuracy and the fraction of remaining frames change as \(th_1\) increases. 
A stricter threshold (larger \(th_1\)) removes more frames and reduces the computational load at the risk of accuracy degradation. We allowed up to 0.3\% reduction in accuracy relative to the baseline configuration without frame skipping. Under these conditions, the three models achieved an optimal balance when \(th_1 = 0.5^6 \cdot \text{max}(\textit{STE})\), reducing the total frame count by 34.42\% on average while keeping accuracy losses within 0.3\%.

\begin{figure}
\vskip -6pt
\centering
\includegraphics[width=0.92\columnwidth]{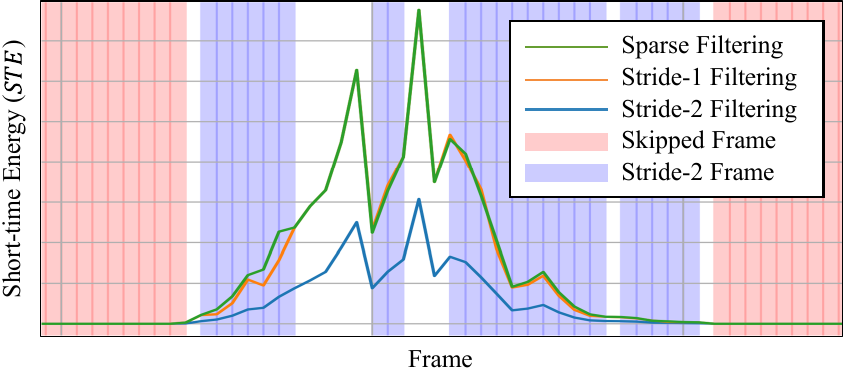}
\vskip -4pt
\caption{Comparison of Short-Time Energy trends: sparsity-aware filtering (labeled ‘sparse filtering’ in the legend) vs. stride-1/2 in a single audio input.}
\label{fig:skip2}
\vskip 3pt
\end{figure}

\begin{figure}
\centering
\includegraphics[width=0.98\columnwidth]{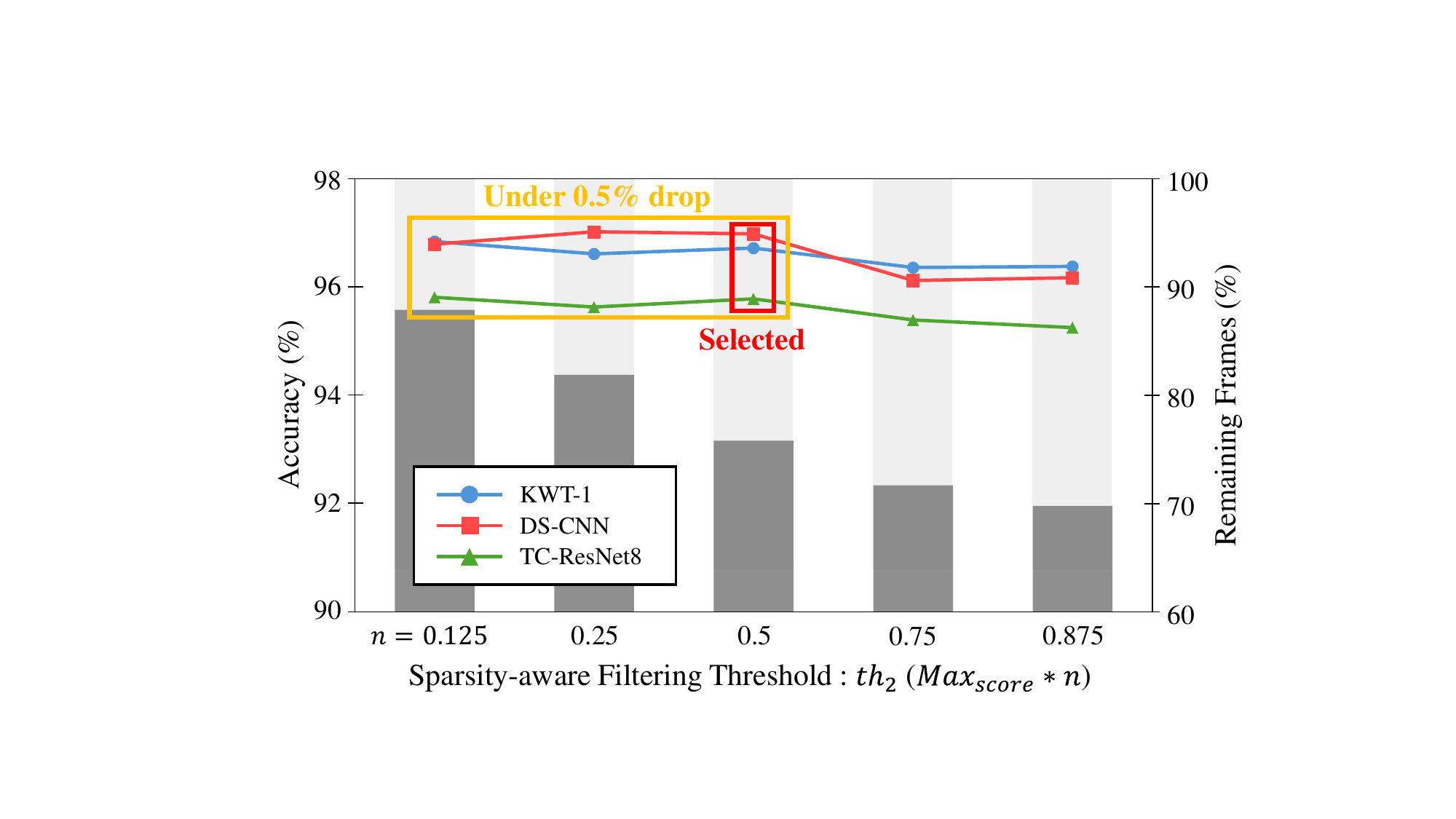}
\vskip -4pt
\caption{Effect of threshold $th_2$ for sparsity-aware filtering on KWS accuracy and remaining frames.}
\label{fig:sparse}
\end{figure}

\subsection{Sparsity-aware Filtering}

Although the frame-skipping method effectively removes frames dominated by noise-level signals, multi-channel KWS systems can still face a substantial processing load. 
To address this issue at finer granularity, we introduce a \emph{Sparsity-aware Filtering} method that adaptively applies \emph{stride-1} or \emph{stride-2} filtering based on the relative importance of each frame. 
Specifically, stride-1 filtering processes every sample in the frame, whereas stride-2 filtering uses only the even-indexed samples, halving the sampling rate and the data volume.

We quantify the importance of a frame for KWS through a score \(S_{\text{frame}}\), which combines two indicators of the speech activity: zero crossing count (\textit{ZCC}) and amplitude difference \(\,A_{\text{diff}}\). 
The \textit{ZCC} is defined as

\vspace{-6pt}
\begingroup
\footnotesize
\begin{equation}
\label{eq:eq3}
ZCC \;=\; \sum_{n=1}^{N-1} \bigl[\text{sign}(x[n])\,\cdot\,\text{sign}(x[n-1]) < 0\bigr],
\end{equation}
\endgroup
where each term corresponds to a sign change between consecutive samples \(x[n]\). 
The amplitude difference $A_{\text{diff}}$, expressed as

\vspace{-2pt}
\begingroup
\footnotesize
\begin{equation}
\label{eq:eq4}
A_{\text{diff}} \;=\; \max\bigl(x[n]\bigr)\;-\;\min\bigl(x[n]\bigr), \quad n \in \text{frame},
\end{equation}
\endgroup
captures the variation in sample amplitudes within the frame. 
We then compute $S_{\text{frame}}$ as follows:

\vspace{-2pt}
\begingroup
\footnotesize
\begin{equation}
\label{eq:eq5}
S_{\text{frame}} \;=\; \alpha\,\cdot\,ZCC \;+\;\beta\,\cdot\,A_{\text{diff}},
\end{equation}
\endgroup
where \(\alpha\) and \(\beta\) modulate each feature’s influence; empirical results suggest that \(\alpha=1\) and \(\beta=0.5\) best capture a frame’s relevance to KWS.

Once \(S_{\text{frame}}\) is calculated, a threshold \(th_2\) determines whether the frame will be subjected to stride-2 (when \(S_{\text{frame}} < th_2\)) or stride-1 filtering (when \(S_{\text{frame}} \geq th_2\)). 
Frames marked for stride-2 downsampling use a low-pass filtered raw waveform to avoid aliasing at the reduced Nyquist frequency; frames deemed important employ stride-1 filtering with a pre-emphasized waveform. 
Because stride-2 filtering effectively halves the sample count, it attenuates the short-time energy (\textit{STE}) compared to stride-1. 
To correct this, we boost the stride-2 \textit{STE} to match the amplitude scale observed under stride-1, leveraging a factor derived from adjacent stride-1 frames. 
As illustrated in~\refFigure{fig:skip2}, the output from Sparsity-aware Filtering realigns the downsampled frame’s \textit{STE} curve with the stride-1 reference.

We further explored how varying \(th_2\) impacts model accuracy and data reduction, evaluating three KWS models (TC-ResNet8, DS-CNN, and KWT-1) under threshold values of \(\{0.125, 0.25, 0.5, 0.75, 0.875\}\) times the maximum \(S_{\text{frame}}\) for each audio input. 
As shown in~\refFigure{fig:sparse}, raising \(th_2\) forces more frames to undergo stride-2 filtering, thereby lowering the remaining frame count at the expense of a slight accuracy drop. 
Constraining accuracy degradation to 0.5\% below the baseline, we found the optimal point at \(th_2=0.5 \cdot \max(S_{\text{frame}})\), reducing per-frame data by 24.2\%. 
Combined with a 34.4\% reduction from frame skipping, the overall Sparsity-aware Data Reduction yields a total 50.3\% cut in data volume. 
In addition, incorporating half-overlapped framing, which provides an additional 24.9\% reduction, brings the total data reduction to approximately 62.7\%.

\section{Dynamic Parallel Processing}
\label{sec:dpp}

\begingroup
\DontPrintSemicolon
\begin{algorithm2e}[t]
    \caption{Dynamic Parallel Processing Algorithm}
    \label{alg:dpp}
    \footnotesize

\KwIn{$N$, $m_{max}$}
\KwOut{$m_{min}$, $m_{opt}$}

\SetKwFunction{FComputeNN}{Find\_m\_MIN}
\SetKwProg{Fn}{Function}{:}{}
\Fn{\FComputeNN{$N$, $m_{\max}$}}{
  $i \gets 1$; \While{$i \le m_{\max}$}{
    \If{$latency(i)\le \frac{T_{audio}}{N}$}{ $m_{\min}\gets i$; \textbf{break} }
    $i \gets i+1$;
  } \Return $m_{\min}$
}

\SetKwFunction{FComputeNOpt}{Find\_m\_OPT}
\SetKwProg{Fn}{Function}{:}{}
\Fn{\FComputeNOpt{$m_{\min}$, $m_{\max}$}}{
  $j \gets m_{\min}+1$; \While{$j \le m_{\max}$}{
    \If{$energy(j) > energy(j-1)$}{ $m_{opt}\gets j-1$; \textbf{break} }
    $j \gets j+1$;
  } \Return $m_{opt}$
}

%

\end{algorithm2e}
\endgroup

\DontPrintSemicolon
\begin{algorithm2e}[t]
    \caption{Filter Scheduling Algorithm}
    \label{alg:H-DBSCAN}

\footnotesize

STRIDES : Stride array for all Frames \;

\SetKwFunction{FAssignPriorities}{ASSIGN\_PRIORITIES}
\SetKwProg{Fn}{Function}{:}{}
\Fn{\FAssignPriorities{STRIDES}}{

    \For{all $s_{current}$ $\in$ STRIDES}{

        \If{$s_{current} = 0$}{
            Apply $Frame Skipping$ to current Frame \;
        }
        \ElseIf{$s_{current} = 1$}{
            \If{$s_{next} = 2$ \textbf{or} $s_{prev} = 2$}{ 
                Priority1.add(STRIDE2, [$Frame\_Index$]) \;
                Priority1.add(STRIDE1, [$Frame\_Index$]) \;
            }
            \Else{
                Priority3.add(STRIDE1, [$Frame\_Index$]) \;
            }
        }
        \ElseIf{$s_{current} = 2$}{
            Priority2.add(STRIDE2, [$Frame\_Index$]) \;
        }
    }
Priority\_Queue $\gets$ stack(Priority1, Priority2, Priority3) \;
    \Return Priority\_Queue \;
}

\SetKwFunction{FScheduling}{SCHEDULING}
\SetKwProg{Fn}{Procedure}{:}{}
\Fn{\FScheduling{STRIDES}}{

    Priority\_Queue $\gets$ ASSIGN\_PRIORITIES(STRIDES) \;

    \While{there are remaining tasks}{
        \ForEach{filter in FILTERS}{
            \If{filter is available}{
                task $\gets$ GET\_NEXT\_TASK(Priority\_Queue) \;
                ASSIGN\_TASK(filter, task) 
            }
        }
    }
}
\end{algorithm2e}

Although the proposed Half-overlapped IIR Framing and Sparsity-aware Data Reduction techniques achieve an average 62.73\% reduction in computational load, this alone cannot support KWS with three or more channels in real-time. 
Therefore, we introduce a parallelization approach that leverages the characteristics of IIR Filtering to ensure adequate performance in multi-channel settings.

Recall that in an IIR Filtering-based FE, nearly every operation except the \textit{log2} transform (e.g., Pre-emphasis, IIR stride-1/stride-2 BPF, and IIR LPF for stride-2) relies on repetitive combinations of the same filter structure. 
Consequently, deploying multiple filter modules in parallel can perform identical operations simultaneously and drastically improve throughput. 
However, edge processors face limited resources and power budgets, preventing an unbounded increase in the number of filter modules $m$. 
Although increasing $m$ generally shortens processing time (latency), it also increases resource usage and power consumption roughly linearly. 
Furthermore, distributing a fixed workload over many modules reduces the marginal latency benefit per additional filter, resulting in a convex profile for energy consumption and implying the presence of a minimum energy point.

\begin{figure}
    \centering
    \vskip -1pt
    \includegraphics[width=0.8\columnwidth]{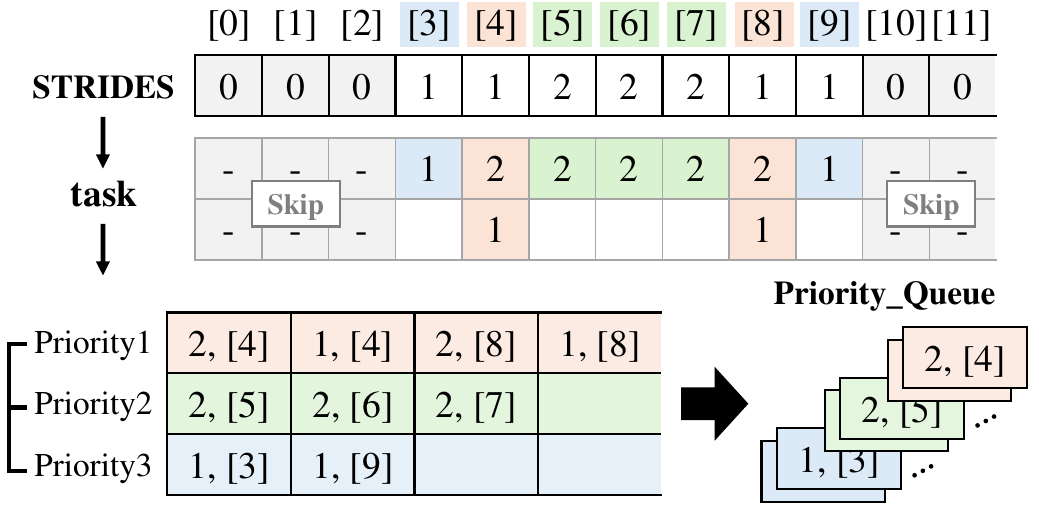}
    \vskip -4pt
    \caption{Procedure for generating the Priority\_Queue.}
    \label{fig:pseudo}
\end{figure}

\begin{figure*}
\centering
\includegraphics[width=2\columnwidth]{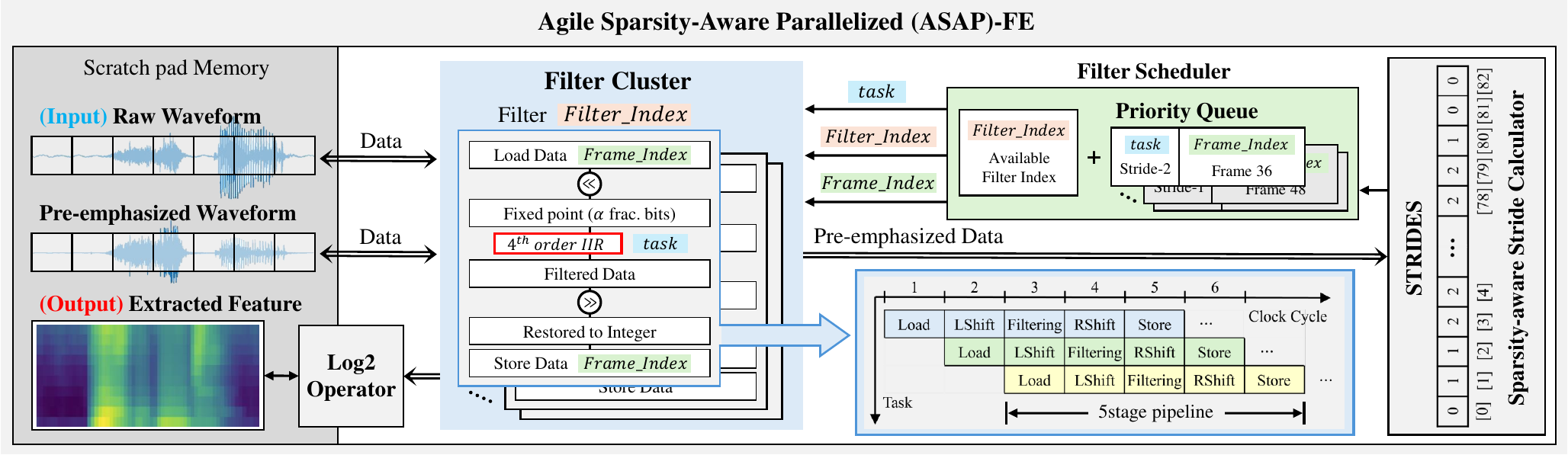}
\vskip -2pt
\caption{Block Diagram of the ASAP-FE Architecture.}
\label{fig:imp}
\vskip -6pt
\end{figure*}

To address these concerns, we decompose the parallelization problem for $N$-channel KWS into two primary subproblems:
\begin{enumerate}
    \item \textbf{P1:} Determine the minimum number of filter modules, $m_{\min}$, required to accommodate $N$ channels:
    \[
        \textit{latency}(m_{\min}) \,\leq\, \frac{T_{audio}}{N}, 
        \quad 
        m_{\min} \,\leq\, m_{\max}.
    \]
    Here, $\textit{latency}(m)$ is the time for an FE-dedicated HW (accelerator) with $m$ filter modules to complete one FE pass, and $m_{\max}$ is the maximum number of modules allowed by the power budget.

    \item \textbf{P2:} Identify $m_{\text{opt}}$ in the range $m_{\min} \leq m \leq m_{\max}$ that minimizes energy consumption:
    \[
        m_{\text{opt}} = \underset{m \,\in\,[m_{\min},\,m_{\max}]}{\mathrm{arg\,min}} \, \bigl(\,\textit{energy}(m)\bigr).
    \]
    The function $\textit{energy}(m)$ quantifies the energy consumed by $m$ filter modules in processing one FE pass. We assume that the FE accelerator can be power-gated during the idle interval $\frac{T_{audio}}{N} - \textit{latency}(m)$, thus reducing overall system energy.
\end{enumerate}

To solve the above sub-problems, we propose the {Dynamic Parallel Processing (\emph{DPP})} approach. \refAlgo{alg:dpp} presents the complete DPP framework in pseudo code, encompassing \texttt{Find\_m\_MIN} for P1 and \texttt{Find\_m\_OPT} for P2. 
The key objective lies in determining \textit{latency}($m$) and \textit{energy}($m$) for a given $m$, which depends on how the filters are scheduled and executed in parallel.

We schedule all filters so that no module remains idle, thus achieving the minimum {latency} at each $m$ and the minimum corresponding {energy} requirement. 
To that end, \refAlgo{alg:H-DBSCAN} details our \emph{Filter Scheduling Algorithm}, which consists of \texttt{ASSIGN\_PRIORITIES} and \texttt{SCHEDULING}. 
This algorithm adopts a \emph{priority-based round-robin} scheme to distribute tasks across filter modules, preventing bottlenecks caused by uneven workloads and ultimately minimizing {latency}. 
Further details of these functions are explained as follows:
\vspace {2pt}

\noindent
\textbf{ASSIGN\_PRIORITIES:}
We schedule only the two core operations, IIR stride-1 and stride-2 BPF, because Pre-emphasis runs exactly once, and LPF is invoked preemptively only for stride-2 processing (thus not requiring explicit priority). 
\texttt{ASSIGN\_PRIORITIES} receives the \emph{stride} array from Sparsity-aware Data Reduction and assigns a priority to each frame. As shown in \refFigure{fig:pseudo}:
\begin{itemize}[leftmargin=1.5em]
    \item Frames with \textit{stride} $=0$ are skipped (Frame Skipping).
    \item Frames with \textit{stride} $=1$ are checked for neighboring frames where \textit{stride} $=2$. If such a neighbor exists, both \textit{stride-1} and \textit{stride-2} operations must occur in the same frame, so we assign it the highest priority (Priority 1). Otherwise, it falls into Priority 3.
    \item Frames with \textit{stride} $=2$ are assigned Priority 2.
\end{itemize}
These prioritized tasks are collected into \texttt{Priority1}, \texttt{Priority2}, and \texttt{Priority3} stacks before being combined into a single \texttt{Priority\_Queue}.

\vspace {2pt}
\noindent
\textbf{SCHEDULING:}
\texttt{SCHEDULING} repeatedly fetches the highest-priority tasks from \texttt{Priority\_Queue} (\emph{priority-based round-robin}) and dispatches them to available filter modules. 
This design ensures that no filter remains idle and that no single module becomes a bottleneck. 
For example, a stride-1 frame adjacent to a stride-2 region necessitates two operations, so it obtains Priority 1 and must be completed before the corresponding stride-2 frame can finish. 
By contrast, a standalone stride-1 frame (requiring no correction) is assigned Priority 3 and can be processed independently.

We achieve minimal latency under each filter configuration by repeatedly invoking \texttt{ASSIGN\_TASK}. 
Subsequently, implementing both the \emph{Filter Scheduler} (which executes the Filter Scheduling Algorithm) and the \emph{Filter Cluster} with $m$ filters in hardware enables us to empirically measure \textit{latency}($m$) and \textit{energy}($m$) for \refAlgo{alg:dpp}. 
In the following Experimental Work section, we present real hardware measurement results and analyze how DPP improves the performance and energy efficiency of $N$-channel KWS processors.

\section{Experimental Work}
\subsection{ASAP-FE Hardware Implementation}
We incorporate the three proposed methods: half-overlapped IIR fragmenting, sparsity-aware data reduction, and dynamic parallel processing into a dedicated hardware system, ASAP-FE. 
As illustrated in \refFigure{fig:imp}, ASAP-FE uses a scratchpad memory (SPM) to store and retrieve raw waveforms, pre-emphasized waveforms, and the final extracted features. 
These data are then processed by the Filter Cluster, where each filter operates internally with integer input/output and fixed-point arithmetic, avoiding the overhead of floating-point operations. 
A five-stage pipeline in the Filter Cluster enables high-throughput processing for tasks such as Pre-emphasis, IIR stride-1/stride-2 bandpass filtering, and IIR low-pass filtering for stride-2.

During Pre-emphasis, a Sparsity-aware Stride Calculator determines the stride pattern for each frame, passing these values to a Filter Scheduler. 
The scheduler uses these stride values to assign tasks, including the relevant filtering operations and frame indices, to a \texttt{Priority\_Queue}, then distributes them to available filter modules in a parallelized manner. 
Once filtering is complete, the energy value of each frame is passed through a log2 operator based on LUT to produce the final extracted feature.

\begin{figure}
\vskip -8pt
\centering
\includegraphics[width=1\columnwidth]{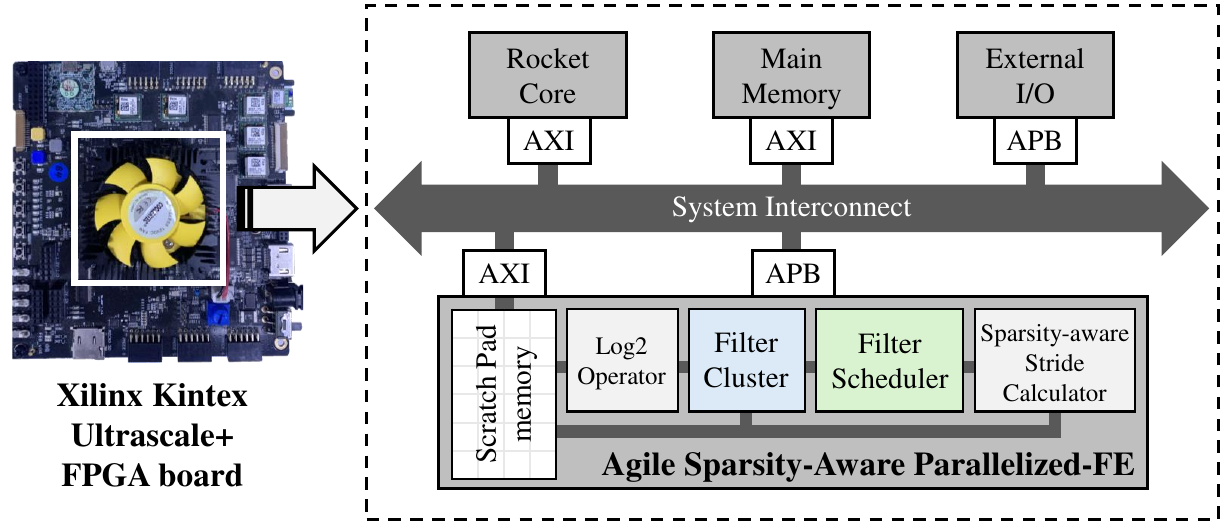}
\vskip -2pt
\caption{Block diagram of the developed processor architecture with ASAP-FE, prototyped on an FPGA.}
\label{fig:arc}
\vskip 4pt
\end{figure}

For RTL design, we developed ASAP-FE and integrated it into an edge processor, parameterizing the Filter Cluster so that the number of filters $m$ can be varied as needed. 
\refFigure{fig:arc} shows the processor architecture designed using {RISC-V eXpress} (\textit{RVX})~\cite{Han:IoT2021,Jang:ACCESS21}. It includes a RISC-V Rocket core~\cite{Rocket} running at 50\,MHz (well established clock speed for edge processor\cite{Park:DATE23, Choi:JSA24, Park:TCAS24}), a 64\,KB on-chip SRAM, and a low-power \(\mu\)NoC interconnect. 
ASAP-FE connects to this processor via an AXI port for SPM transactions and an APB port for configuration and control.

To validate our design, we prototyped multiple processors containing ASAP-FE with $1 \le m \le 30$ on a Xilinx Kintex Ultrascale+ FPGA board~\cite{UltraScale}. We performed logic synthesis at the 45\,nm node using the NCSU 45\,nm PDK library~\cite{NCSU}. 
\refTable{table:resource_power} presents the resource utilization and power consumption measured for the $m=15$ configuration. 
The FPGA evaluation with Xilinx Vivado~\cite{Vivado} confirms that ASAP-FE occupies 22{,}725 LUTs and 16{,}119 flip-flops, which correspond to roughly 40\% of the processor’s LUTs and 30.9\% of its FFs. At the same time, Synopsys Design Compiler~\cite{DesignCompiler} reports a total power of 39.5\,mW for the entire system, of which ASAP-FE constitutes 10.9\,mW (27.6\%). 
Although this overhead is not insignificant, the $m=15$ setup supports real-time KWS for up to 25 channels, a capability that substantially enhances multi-channel performance in edge applications and makes the overhead acceptable in practice. 
The parameterizable architecture allows designers to optimize $m$ for different performance and energy requirements, further broadening the applicability of ASAP-FE in resource-constrained scenarios.

\begin{table}
\vskip -4pt
	\caption{FPGA resource and power consumption breakdown of ASAP-FE applied processor when Filter cluster w/ 15 Filter modules.}
	\centering
\begin{center}


\resizebox{0.9\columnwidth}{!}{
    
\renewcommand{\arraystretch}{1.1}
\begin{tabular}{cll|c|c|c}
\Xhline{1pt}
\multicolumn{3}{c|}{{IPs}}            & {LUTs}  & {FFs} & $P_{total}$\,$(mW)$ \\ \Xhline{1pt}

\multicolumn{3}{c|}{\cellcolor[HTML]{EFEFEF}Developed Processor}  
& \cellcolor[HTML]{EFEFEF}56,772  
& \cellcolor[HTML]{EFEFEF}52,212  
& \cellcolor[HTML]{EFEFEF}39.5 \\ \hline

\multicolumn{3}{l|}{$\llcorner$ RISC-V Rocket Core}  & 14,993  & 9,762  & 10.84 \\ \hline
\multicolumn{3}{l|}{$\llcorner$ Rocket Core Interface} & 159    & 196    & 0.17 \\ \hline
\multicolumn{3}{l|}{$\llcorner$ External I/O}       & 2,813   & 2,133  & 5.38 \\ \hline
\multicolumn{3}{l|}{$\llcorner$ Main Memory}        & 164     & 315    & 1.25 \\ \hline
\multicolumn{3}{l|}{$\llcorner$ System Interconnect} & 15,918  & 23,687 & 10.96 \\ \hline

\multicolumn{3}{l|}{\textbf{$\llcorner$} \textbf{ASAP-FE}}   & \textbf{22,725} & \textbf{16,119} & \textbf{10.9} \\ \hline

\end{tabular}
}
\end{center}
	\label{table:resource_power}
\end{table}

\subsection{Evaluation of ASAP-FE in Multi-Channel KWS}
\label{sec:eval_asap_fe}

\begin{figure}
\vskip -10pt
\centering
\includegraphics[width=1\columnwidth]{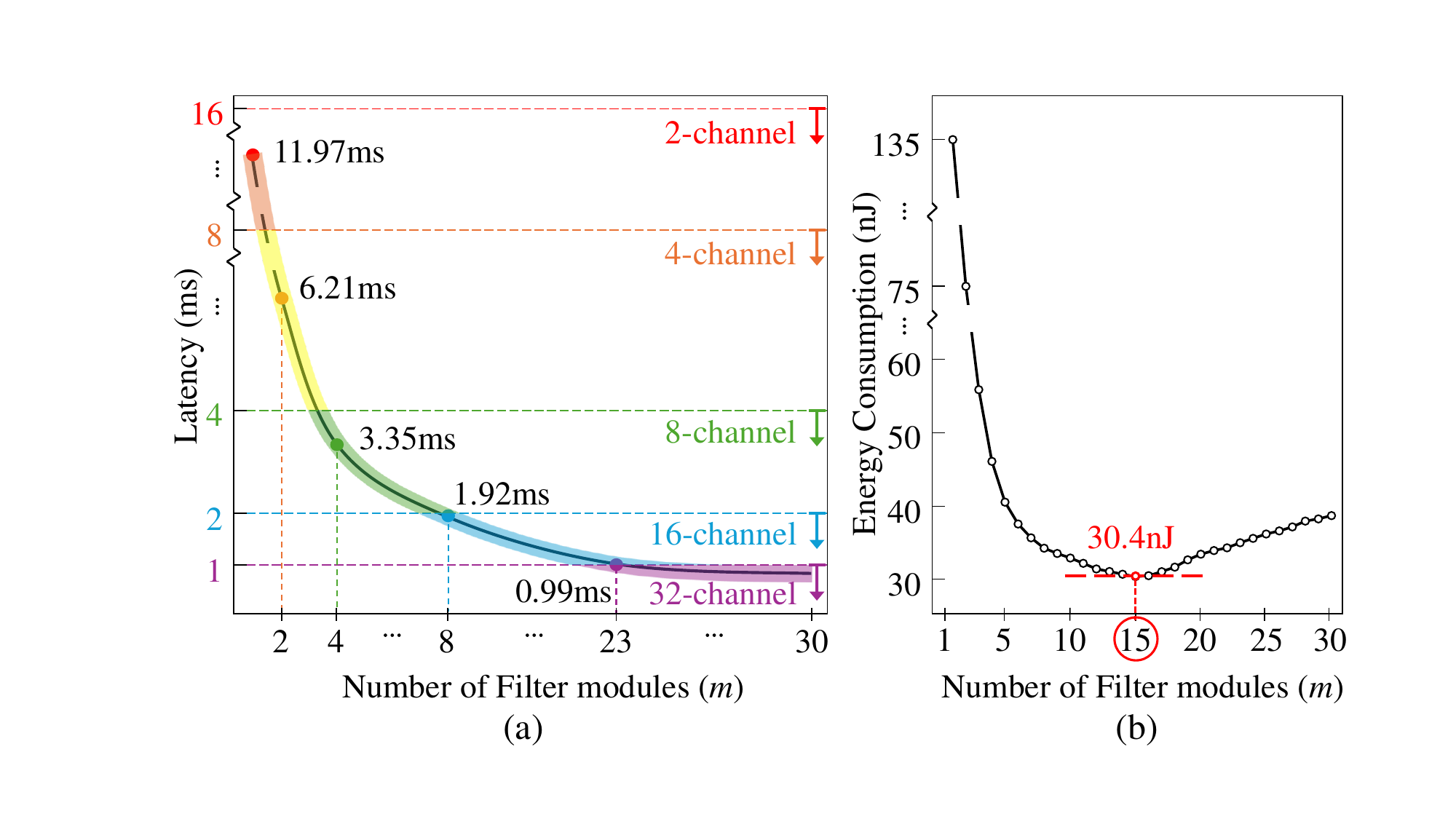}
\vskip -4pt
\caption{Effect of the Number of Filter Modules in Filter Cluster on (a) Latency (b) Energy Consumption.}
\label{fig:optimal}
\vskip 2pt
\end{figure}

To validate the performance of our proposed ASAP-FE, we measured the latency, energy consumption, and recognition accuracy of multi-channel KWS running on various prototype processors.
We used the Google Speech Commands Dataset (GSCD) for all experiments. 
\refFigure{fig:optimal}(a) illustrates how the average latency changes as we increase the number of filter modules in the Filter Cluster ($m$) from 1 to 30, and also identifies the minimum $m_{\min}$ required to support $N$-channel KWS for each $N$.

We first established a baseline by measuring a mono-channel KWS with fully overlapped IIR Filtering method, which required 32\,ms per FE pass. 
We thus set $T_{audio}=32$\,ms. 
Owing to ASAP-FE’s 62.73\% reduction in data, even a single filter ($m=1$) achieves a latency of 11.97\,ms, substantially below the baseline. 
As $m$ increases, latency drops quickly; at $m=30$, latency is only 0.88\,ms. 
Consequently, the number of channels supported in real time rises in tandem with the reduction in latency.  
For example, once the ASAP-FE has at least 23 filters, the system can support up to 32 channels within the same $T_{audio}$.

Next, we evaluated how \emph{DPP} finds an energy-optimal filter configuration $m_{\text{opt}}$ by measuring the total energy consumption of ASAP-FE for various $m$ values, as shown in \refFigure{fig:optimal}(b). 
Although latency decreases sublinearly with $m$, power consumption increases nearly linearly with the number of active filter modules, causing energy consumption to drop initially and climb after passing an optimal point. 
This pattern matches the DPP analysis, which predicts that higher parallelism shortens the execution time, and thus lowers the energy consumption to a threshold beyond which the power overhead dominates, and the overall energy increases again. 
Our results show that $m=15$ yields the minimum energy consumption of 30.4\,nJ, with a latency of 1.25\,ms—sufficient to support real-time KWS up to 25 channels. 
Hence, if energy efficiency is the primary design concern, using 15 filters is optimal for systems requiring up to 25 channels. 
We also confirm that if more than 25 channels are needed, $m_{\min}$ itself becomes the optimal choice ($m_{\text{opt}} = m_{\min}$).

\begin{table}
\vskip -2pt
	\caption{Accuracy results by running TC-ResNet8, DS-CNN, KWT-1}
	\centering
\begin{center}

\renewcommand{\arraystretch}{1.2} 
\resizebox{0.99\columnwidth}{!}{%
\renewcommand{\arraystretch}{1.3}
\begin{tabular}{c|c|c|c}
\Xhline{1pt}
\diagbox[dir=NW,width=4.8cm,height=1.3\line]{}{} & TC-ResNet8 & DS-CNN & KWT-1 \\  
\Xhline{1pt}
Fully overlapped IIR Filtering  & 96.35\%  & 97.13\%  & 97.28\% \\ \hline

\textbf{ASAP-FE}  & \textbf{95.43\%}  & \textbf{96.22\%}  & \textbf{96.48\%} \\ \hline

\end{tabular}%
}

\end{center}
	\label{table:final_acc}
\vskip -2pt
\end{table}

Finally, we verified that ASAP-FE delivers adequate accuracy despite its latency and energy gains by comparing it with the baseline fully overlapped IIR Filtering. 
As reported in Table~\ref{table:final_acc}, we tested the TC-ResNet8, DS-CNN, and KWT-1 models. The baseline method achieves 96.35\%, 97.13\%, and 97.28\% accuracy, respectively, while ASAP-FE scores 95.43\%, 96.22\%, and 96.48\%, representing an overall drop of under 1\%. 
This is well below the commonly accepted threshold of about 1.4\% accuracy loss~\cite{Xiao:ISSCC2024, Xiao:ISCAS2024}, indicating that ASAP-FE remains highly competitive from an accuracy standpoint while significantly reducing both latency and energy.

\vspace {8pt}
\section{Conclusion}
This paper presented ASAP-FE, a hardware-oriented front-end tailored to meet the formidable demands of multi-channel KWS in edge environments, an area largely unaddressed by existing single-channel solutions.
The proposed design integrates three key techniques: Half-overlapped IIR Framing, Sparsity-aware Data Reduction, and Dynamic Parallel Processing to achieve a notable 62.7\% cut in front-end workload while maintaining accuracy losses under 1\%. 
Specifically, the half-overlapped framing technique preserves crucial speech transitions, yet reduces redundant filtering by about 25\%. 
The sparsity-aware data reduction effectively skips noise-dominated frames and selectively applies stride-2 filtering to less significant frames, reducing data volume by an additional 50\%. 
Dynamic parallel processing further enhances throughput by distributing IIR filtering tasks to multiple filter modules under a priority-based scheduler, balancing latency and energy consumption.

We implemented ASAP-FE-equipped edge processors in Verilog RTL and demonstrated their viability through FPGA prototyping on Xilinx Kintex Ultrascale+ boards, then synthesized the designs at a 45 nm node. 
Experimental results show that ASAP-FE scales to 32-channel KWS within a 50 MHz operational budget, supporting real-time operation with minimal power overhead. 
Moreover, we identified an energy-optimal point of 15 filter modules for systems requiring up to 25 channels, reducing feature-extraction latency to 1.25 ms and total energy usage to 30.4 nJ per pass. 
These findings suggest that ASAP-FE accommodates demanding multi-channel scenarios and mitigates the stringent power constraints typical of edge devices. 
The parameterizable filter cluster allows system architects to tailor the design for diverse workloads, further broadening ASAP-FE’s applicability to advanced IoT, surveillance, and other audio-intensive applications. 
Overall, ASAP-FE effectively bridges the gap between real-time performance and low-power requirements in emerging multi-channel KWS systems.

\newpage

\bibliographystyle{IEEEtran}
\bibliography{reference}

\end{document}